\documentclass[a4paper,12pt]{article}
\usepackage[utf8]{inputenc}
\usepackage{titlesec}
\usepackage[margin=2.5cm]{geometry}
\usepackage{amsmath, amsfonts}
\usepackage{verbatim}
\usepackage[numbers]{natbib}
\usepackage[title,titletoc]{appendix}
\usepackage{graphicx}
\usepackage{epstopdf}
\usepackage{xcolor}
\usepackage{array}
\usepackage{hyperref}
\usepackage{hyperref}
\usepackage{lmodern}
\usepackage{hyperref}
\usepackage{lmodern}

\DeclareGraphicsExtensions{.eps}
\usepackage{subcaption}

\begin{document}

\title{Technology networks: the autocatalytic origins of innovation}

\author{Lorenzo Napolitano$^\textrm{1,3}$, Evangelos Evangelou$^\textrm{1}$, Emanuele Pugliese$^\textrm{1,2,3}$,\\ Paolo Zeppini$^\textrm{1,4}$ and Graham Room$^\textrm{1}$.}

\maketitle
\thispagestyle{empty}
\small
\begin{center}
    $^\textrm{1}$ \emph{University of Bath, BA2 7AY Bath, UK}\\[7pt]
    $^\textrm{2}$ \emph{International Finance Corporation, World Bank Group, 20433 Washington, USA}\footnote{The findings, interpretations and conclusions expressed herein are those of the authors and do not necessarily reflect the view of the World Bank Group, its Board of Directors or the governments they represent.}\\[7pt]
    $^\textrm{3}$ \emph{Istituto dei Sistemi Complessi (ISC)-CNR, 00185 Rome, Italy}\\[7pt]
    $^\textrm{4}$ \emph{Universit\'e C\^ote d'Azur, CNRS, GREDEG, 06560 Valbonne, France}
\end{center}

\begin{abstract}
We analyse the autocatalytic structure of technological networks and evaluate its significance for the dynamics of innovation patenting.
To this aim, we define a directed network of technological fields based on the International Patents Classification, in which a source node is connected to a receiver node via a link if patenting activity in the source field anticipates patents in the receiver field in the same region more frequently than we would expect at random.
We show that the evolution of the technology network is compatible with the presence of a growing autocatalytic structure, \emph{i.e.} a portion of the network in which technological fields mutually benefit from being connected to one another.
We further show that technological fields in the core of the autocatalytic set display greater fitness, \emph{i.e.} they tend to appear in a greater number of patents, thus suggesting the presence of positive spillovers as well as positive reinforcement.
Finally, we observe that \emph{core shifts} take place whereby different groups of technology fields alternate within the autocatalytic structure; this points to the importance of recombinant innovation taking place between close as well as distant fields of the hierarchical classification of technological fields.
\end{abstract}

%--------------------------------------------------------------------------------------------------------------------------------------

\section{Introduction}

A large body of research on complex systems -- physical, biological and socio-economic as well -- has focused on the relation between the structure of interactions within heterogeneous populations of agents and the dynamic properties of the aggregate system they populate.
This has implied a change of perspective from \emph{linear} narratives, in which the direction of causality connecting phenomena is unambiguous, towards processes of cumulative causation \citep{kauffman1986autocatalytic, Arthur:1989}.
This relation is particularly evident in the study of biological systems, where models of pure resource competition are unable to explain the persistent variety of ecosystems.
To this aim, it is useful to introduce the idea that some species in a heterogeneous population can serve as catalysts (or inhibitors) for the survival of other species.
The interaction between preys and predators is one of the best-known examples of this kind of relation, but a similar mechanism has also been observed in settings like plant-pollinator interactions, opportunistic behaviour, and symbiotic relations \cite{dominguez2015ranking}.
A particularly relevant feature of these ecological systems is the presence of \emph{autocatalytic sets} \citep{Jain-Krishna:2002} -- self-sustaining subsystems, in which each species benefits directly or indirectly from its cohabitation with the others.
The relevance of interactions in the above framework lends itself to a complex systems interpretation for which networks are a natural tool of analysis.

The idea of catalytic interaction can be fruitfully extended to human systems and, in particular, to the realm of technological innovation.
In this setting, interactions take place between technological fields whenever existing ideas are applied to new problems or used to bridge previously unrelated fields thereby expanding the set of technological capabilities and spawning further innovation \citep{Schumpeter:1934, Nelson-Winter:1982, Frenken-etal:2012, strumsky2012using, loreto2016dynamics, iacopini2018network}.
Consider, for instance, the development of lasers, which have opened the way to a number of innovations in different industries, ranging from telecommunication to data storage and health care.
It is also possible that the combination of multiple technological fields will give rise to a radical -- possibly disruptive -- novelty or to technological `convergence' \citep{patel1997technological, Gambardella-Torrisi:1998, korzinov2017general}.
An example of the former is the case of opto-electronics, since optical and electrical devices lie at the root of the technological framework for modern telecommunication systems; a notable example of convergence is instead provided by smart-phones and electronic tablets, which combine functionalities that could previously be found only separately in computers, telephones, and television sets.
In general, combining knowledge from previously isolated domains has become extremely relevant in several innovation-oriented domains, such as academic research projects -- which often involve scientific collaborations between groups with heterogeneous backgrounds \citep{Powell-etal:2005} -- and industrial endeavours -- where R\&D collaborations have become common practice \citep{Ahuja:2000, Hagedoorn:2002, konig2012efficiency} especially in sectors characterized by a quick pace of technological progress (\emph{e.g.} biotechnology \citep{Roijakkers-Hagedoorn:2006} and IT \citep{Hanaki-etal:2010}).
For this reason, the network structure of both scientific \citep{newman2001structure, Barabasi-etal2002, Goyal-etal:2006} and industrial collaborations \citep{letterie2008information} has been studied in depth in the past.

This study investigates the autocatalytic structure of the network of interactions between technological fields extracted from patent data and shows how the tools of complex systems analysis are able to forecast the evolution and future relevance of individual fields based on their role in the interaction network.
Following \cite{puglieseetal2017}, we say that a technological field is catalytic for another one if the development in a region of innovations involving the former is positively associated with the future emergence of innovative expertise in the latter in the same region.
We employ patents as a proxy for inventions, in line with an established body of research about the patterns of technological change, which has been pioneered by the scholarly \citep{Griliches:1998, hall2001nber} and institutional effort to tap into their potential to shed light on relevant open questions concerning the drivers of technological progress, the relative importance of technological domains, and the significance of technological proximity \emph{vis \`a vis} technological variety for the emergence of radical and incremental innovation \citep{Castaldi-etal:2015}.
One of the decisive advantages of using patents as sources of data stems from the fact that exclusive commercial rights to an invention are granted to applicants provided that they publish a complete description of the patented invention allowing it to be replicated by others once the exclusive rights expire.
In order to assess innovativeness, patent offices map the claimed novel features of each invention to the technological fields it impacts through a standard classification system and collect the above information into dedicated databases.
This leads both to an extensive coverage of the innovation spectrum and a high degree of standardization that allows large scale analysis.

The rest of paper is organized as follows.
Section \ref{s:enetwork} defines the technological network based on patent classification codes; Section \ref{s:acnetworks} summarizes the features of autocatalytic networks; Section \ref{s:actnetworks} reports the results of our study of the autocatalytic structure of the technology network; Section \ref{s:conclusion} concludes.

\section{The patent network}\label{s:enetwork}

\subsection{Connecting regions and technological fields}\label{s:regsToFields}

Our analysis relies on the patent data contained in PATSTAT \cite{patstatcatalog}, a comprehensive database collecting information about  applications filed at national and regional patent offices all around the world.  
PATSTAT contains several tables linking over fifty million patent applications to information such as the filing date of applications, the patent families\footnote{A patent family is a collection of patent applications addressing the same invention, usually in different patent offices.} they belong to, and their technological content as encoded by the International Patent Classification (IPC) codes attributed to the patent claims by patent office examiners. 
IPC codes define a hierarchical classification consisting of five levels (\emph{sections, classes, subclasses, groups, subgroups}), which includes 8 codes at the coarsest level (sections) and over 70 thousand codes at the bottom of the classification tree. 

We associate patents to the location of their assignees through Orbis, a commercial database of firm level data 
maintained by the Bureau van Dijk, which collects the list of patent PATSTAT identifiers of the applications filed by companies that have been active in patenting at some point in time.
We match the technological codes associated to patent families with the firm-level data so to unambiguously localize firms geographically through their country of residence and their postal code. 
This allows us to construct a geographical matrix attributing IPC codes to regions through the patent portfolios and the geographical locations of the companies\footnote{
Note in passing that companies (or groups of companies) based in multiple locations do not necessarily assign the patents they own to the address of their corporate headquarters, but sometimes attribute inventions to one of their regional offices or a subsidiary.
}.

We observe different regions worldwide across time to uncover the effects that innovation in a field produces on other fields within the same region.
In order to build comparable regions across different countries, we need a spatial identification system.
For European Countries we connect postal codes of patenting companies to the associated NUTS 3\footnote{Third level of the hierarchical \emph{Nomenclature des Unit\'es Territoriales Statistiques}, the standard classification of European sub-national regions established by the European statistical office. The acronym NUTS is also commonly associated to the English name of the classification, \emph{Nomenclature of Territorial Units for Statistics}.} regions of the standard European classification (corresponding \emph{e.g.} to Provinces in Italy and Districts in the United Kingdom). 
Since extra-European countries are not included in the NUTS classification, we resort to national classifications when necessary and align them to achieve a broad overall accordance with the employed European classification and hence assure the consistency of the geographical tree. 
Our data define a spacial hierarchy comprising over 3000 regions in 39 countries and aggregating information about the location of around 500 thousand patenting firms.

As mentioned above, the basic units of observation for the construction of the data matrices ($\mathbf{W}(t)$) are individual firms and the patents they own. 
In particular, we group patents into families
and consider the latter as individual inventions because of the strong contiguity between documents they group together.
In building the matrices, we assume that every family containing patents filed in year $t$ counts as one unit and weighs accordingly within $\mathbf{W}(t)$. 
Moreover, we make the hypothesis that the technologies expressed within a patent family can be reasonably accounted for by considering the set of unique IPC codes they contain, thus avoiding to double-count codes that appear in patent applications filed in the same year that belong to the same family.
We evenly split the unit of weight attributed to each family that is active in a given year between all unique combinations of technology codes and regions it maps to, therefore defining element $W_{r,i}(t)$ as the sum, for all families containing patents filed in year $t$, of the shares of each family attributed to field $i$ and region $r$.

For the purpose of the analysis, we need to transform each $\mathbf{W}(t)$ into a presence-absence matrix.
In line with the literature, we assign a value of $1$ to a location-technology pair if the value within the corresponding cell is compatible with a measure of revealed advantage. 
This allows us to reduce noise and avoid overstating the relevance of technological fields in those regions in which they play only a marginal role. 
In particular, we use revealed comparative advantage \citep{balassa1965trade} to produce a matrix $\mathbf{M}(t)$ in which $M_{r,i}(t)$ is recorded as a presence ($M_{r,i}(t)=1$) if in the corresponding $\mathbf{W}(t)$ we have that
\begin{equation}
 \frac{W_{r,i}}{\sum_{i} W_{r,i}} > \frac{\sum_{r} W_{r,i}}{\sum_{r,i} W_{r,i}} \mbox{ ,}
 \end{equation}
and an absence ($M_{r,i}(t)=0$) otherwise.  

Due to time lags between patent filing and data publication by patent offices, the version of PATSTAT we employ (2014a) contains reliable data up to 2011, after which coverage falls sharply.
For this reason, 2011 is the most recent year we include in the analysis. 
As for the left extreme of the time interval, coverage is not an issue especially for the second half of the $20^{th}$ century, even though the number of patents filed decreases quickly going backwards. 
We stop at 1980 because it strikes a balance between having a long time interval for the analysis and dealing with yearly $\mathbf{M}(t)$ matrices that are not excessively sparse.

\subsection{Directed network between technological fields}

The aim of this work is to measure the relationship between the patenting activity taking place within a geographical region ($r$) involving technological field $i$ at time $t$ and the patenting activity performed in $r$ involving a possibly different field ($j$) at time $t+\delta$\footnote{The results presented in this paper refer to $\delta=1$}.
to this end, we count how often patents in field $i$ are present at time $t$ in regions that produce patents in field $j$ at time $t+\delta$.
We discount for regional diversification $d_r(t)$ -- \emph{i.e.} the number of fields in which region $r$ is active at time $t$ -- and the ubiquity of different fields $u_i(t)$ -- \emph{i.e.} the number of regions in which each field is represented at time $t$-- to establish a measure of the excess probability that innovation in a technological field precedes in another field in the same place.
Applying the procedure proposed by \cite{puglieseetal2017} to the conceptual framework proposed by \cite{teece1994understanding, zaccaria2014taxonomy, yan2017measuring}, we obtain:
\begin{equation}
   B_{i,j}(t,\delta)= \frac{1}{u_i(t)}\sum_r\frac{M_{r,i}(t) 
M_{r,j}(t+\delta)}{d_r(t+\delta)} \mbox{ ,}\label{eq:patSpace}
  \end{equation}
where $d_{r}(t)=\sum_{j} M_{r,j}(t)$ and $u_{i}(t)=\sum_{r} M_{r,i}(t)$.
$B_{i,j}(t,\delta)$ can be interpreted as the probability that a region displaying a revealed competitive advantage at time $t$ in the field $i$ will reveal a competitive advantage at time $t+\delta$ in field $j$
\begin{equation}
  B_{i,j}(t,\delta)=Probability(j,t+\delta|i,t)=\sum_r Probability(j,t+\delta|r) 
Probability(r|i,t) \mbox{ ,}\label{eq:probability}
\end{equation}
assuming that the information about the capabilities linking pairs of technological fields is fully captured by their co-occurrence within each region, \emph{i.e.} that $Probability(j,t+\delta|r,i,t)=Probability(j,t+\delta|r)$.

\begin{figure}
 \begin{center}
  \includegraphics{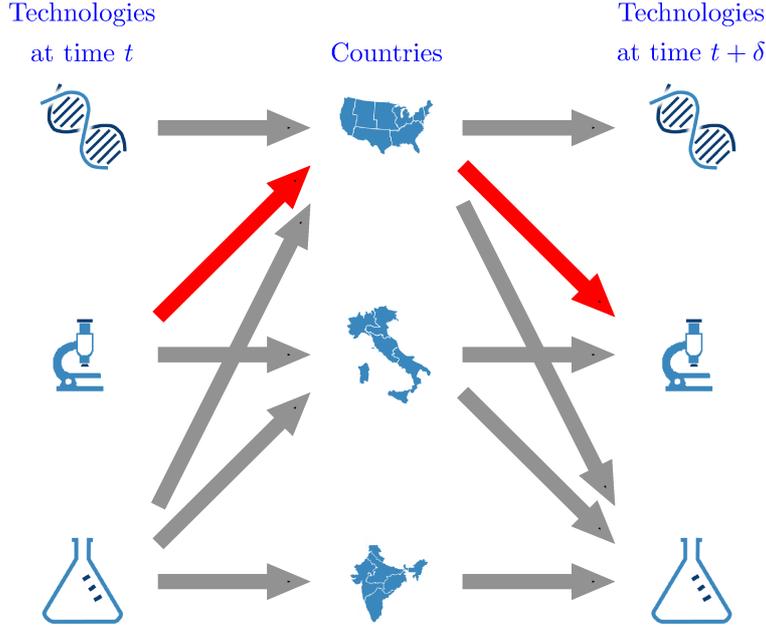}
 \end{center}
 \caption{Model representation of the three-layer technology-country-technology.}
 \label{f:multilayer}
\end{figure}

An equivalent way of interpreting equation \ref{eq:probability} is illustrated in Figure \ref{f:multilayer}, which depicts the tripartite directed network connecting 
 (1) technological field $i$ at time $t$ to
 (2) regions, and regions to
 (3) technological field $j$ at time $t+\delta$.
In this framework, $B_{i,j}(t,\delta)$ is equivalent to the probability that a random walk on the network starting from technology $i$ at time $t$ will reach technology $j$ by time $t+\delta$.
It is worth noting that the choice of the scale of analysis is not neutral, since -- like any socio-economic process -- also innovation has different characteristics depending on the resolution at which it is observed.
However, equation \ref{eq:probability} can be naturally applied to observe the system at different technological granularities.
In what follows, we employ IPC classes for the analysis presented in the main text and later check the robustness of the results on the more disaggregated IPC subclasses (see Supplementary Information). 
Classes and subclasses are two nested levels of the IPC hierarchy that divide the spectrum of technological fields associated to patents respectively into 121 and 640 fields.

\subsection{Assessing link significance}\label{s:link_sig}

Matrix $\mathbf{B}(t,\delta)$ of equation \ref{eq:patSpace} represents a directed weighted network connecting all technologies\footnote{
  For ease of presentation, in the following we omit the time indexes and refer to $\mathbf{B}(t,\delta)$ and its elements as $\mathbf{B}$ and $B_{i,j}$ respectively. 
  The same simplification applies to all the matrices derived from $\mathbf{B}$.
}.
However, to decide whether a link between two technologies is statistically significant, a \emph{null model} is required to account for the fact that some links could seem relevant due to the properties of the graph without, however, being the product of any actual catalytic effect.
For example, very advanced technologies can be developed only by a minority of regions around the World and, for this reason, the associated codes might often appear in the same region at different points in time though no catalytic relation connects them.
Following \cite{puglieseetal2017}, we use the bipartite configuration model \citep{saracco2015randomizing} to test the empirical network against a randomly generated counterpart that displays, on average, the same degree distribution. 
Thus, in the random graphs we generate, each region has the same expected diversification in terms
of technological codes as the empirical data  and each technological code has the empirically observed expected ubiquity.
The degree is thus the only information we extract from the empirical matrices to generate the null models.
Generating a large number of null matrices with the same null model (in our case, 1000 null matrices for each pair of years), it is possible to establish the significance of each link between technologies and generate a matrix $\mathbf{P}$ with element $P_{i,j}$ representing the percentile of the null distribution containing the link associated to $B_{i,j}$.
We can then define the statistical significance of each individual link. 

The final step consists in constructing the unweighed directed adjacency matrix $\mathbf{C}$ that contains only the significant links in the network.
A link from field $i$ to field $j$ is included in $\mathbf{C}$ if the corresponding $P_{i,j}$ is larger than a fixed 
threshold, say $1-p$.
However, if this comparison is performed separately for each pair of fields by looking at individual significance levels, by definition, we would expect a share $p$ of false positive links to be retained. Effectively, the probability of false positive links for the whole network will be higher than the desired level $p$. This is known as the multiple comparisons problem in the statistics literature \citep{colquhoun2014investigation}. A method to accurately control for the proportion of false positive links simultaneously for all pairwise link tests, and thus maintaining the overall significance to $p$, is the false-discovery-rate procedure of \cite{benjamini1995controlling}. The network used in our analysis was constructed using this procedure by retaining significant links in the constructed network from the matrix $\mathbf{P}$ with overall significance level $p=5\%$.

\section{Autocatalytic networks}\label{s:acnetworks}

The matrix $\mathbf{C}$ defined in the Section \ref{s:link_sig} is a directed adjacency matrix representing the links between the
$N$ nodes that represent technological fields: a directed link from a technological field $i$ (source)
to another technological field $j$ (receiver) is present ($C_{i,j}=1$) if there is a significant signal suggesting that
patenting activity in $i$ promoted patenting activity in $j$ in the same region.
Within this representation, directed technological links either exist or not, and the technology network 
is completely specified by a binary \emph{adjacency matrix} of size $N \times N$ with elements $C_{i,j}$.

Inspired by \cite{kauffman1986autocatalytic}, \cite{Jain-Krishna:2001, Jain-Krishna:2002} have proposed a model of network 
evolution for biological systems based on a catalytic interpretation of the relationship between species, which we apply to the case of interactions between technologies. 
To describe the fundamental analytical properties of catalytic systems, we use the simplest dynamical model
\begin{equation}\dot{y}_i=\sum_j C_{i,j} y_j,\label{eq:model}\end{equation}
where $y_i$ is the intensity of activity in technological class $i$. 
In this simple example, activity in field $i$ pushes the innovation activities in every field $j$ such that $C_{i,j}=1$.
Notice that the class of models proposed by \cite{Jain-Krishna:2001} is vast and the following results are still valid for 
more realistic models of the dynamic behaviour of the innovation system.

Consider for instance an empty network ($C_{i,j}=0 \mbox{ } \forall i,j$).
In this toy example, $\dot{y}_i=0 \mbox{ } \forall i$ and any initial condition is an equilibrium.
Let us now add a link from $i$ to $j$. 
In the model, while all the other nodes still experience a constant activity, 
we have $\dot{y}_j=y_i$, which implies $y_j(t)= y_j(0)+y_i(0)t$.
When more links are present, we can have different polynomial behaviours for the activity in different technology classes.
However, an interesting case arises when there is a direct or indirect reciprocal influence between $i$ and $j$.
In this case, $i$ and $j$ form an autocatalytic set, which, in terms of the technology network implies that the more patents employ technology $i$, the
more will also employ technology $j$, and \emph{vice versa}.
The catalytic cycle creates an exponential dynamics involving the innovation activities acting on fields in the set.
This exponential behaviour is in line with the empirical observation of the innovation system, both in 
terms of patents and in terms of productivity growth.
More in general, we can see the same exponential dynamics for any arbitrarily long cycle, \emph{i.e.} for any closed path connecting a subset of the nodes in the network.

\begin{figure}
 \centering
 \includegraphics{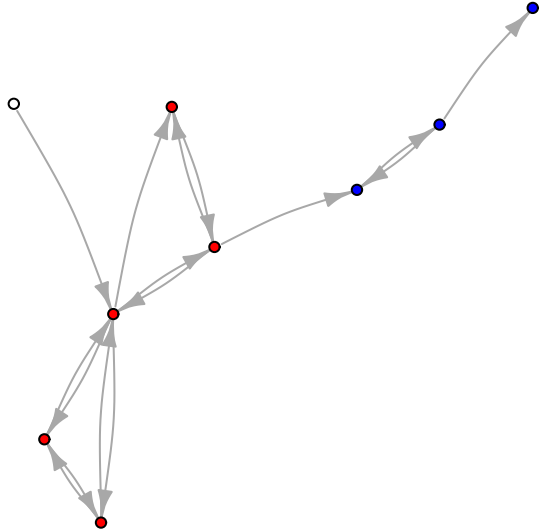}
 \caption{Example of a graph containing an \emph{autocatalytic set} (ACS). The red nodes belong to the \emph{core} of the ACS, the blue nodes belong to the \emph{periphery} of the ACS, while the white node does not belong to the ACS, because it has no incoming link from the ACS itself.}\label{f:example_acs}
\end{figure}

To this end, it is useful to introduce the notion of an \emph{autocatalytic set} (ACS), which is defined as a ``subgraph, each of whose nodes has at least one incoming link from a node belonging to the same subgraph'' \citep{jain2002graph}.
In what follows, any set of nodes connected through one or more cycles will be called the \emph{core} of its corresponding ACS, which, in addition to the nodes in the core, also includes the \emph{periphery}, \emph{i.e} the set of nodes that are catalysed by the core but have no outgoing links feeding into the closed path. 
Because of this configuration, peripheral nodes have a passive role in the ACS and thus are not a part of the core; however, they still benefit from the boost provided by their incoming links.
Figure \ref{f:example_acs} depicts a simple network containing an ACS and highlights the relevant subsets in which its nodes can be partitioned.
By expressing equation \ref{eq:model} in its matrix form, $\mathbf{\dot{y}}=\mathbf{C}\mathbf{y}$, it is easy to show the relationship
between the presence of cycles in the network and the presence of positive \emph{eigenvalues} of the adjacency matrix $\mathbf{C}$. 
Since the adjacency matrix is \emph {non-negative}, 
the Perron-Frobenius theorem guarantees the existence of a real eigenvalue which is the largest of all other 
eigenvalues. This one is called the Perron-Frobenius eigenvalue (PFe), $\lambda_1$. 
It is possible to prove 
\citep{rothblum1975algebraic, Jain-Krishna:2006}
that a cycle exists in the graph if the PFe is greater than 0.
A formal proof of the theorem is beyond the scope of this paper, but it can be shown that, if the PFe is greater than 0, any innovation activity corresponding to a positive element of the Perron-Frobenius Eigenvector (PFE), $\mathbf{y_{1}}$, experiences an exponential growth because $\mathbf{\dot{y}_{1}}=\lambda_1 \mathbf{y_{1}}$.
Indeed, the PFe is informative of the presence or absence of closed directed paths (loops) 
in the graph, and its corresponding Eigenvector has elements different from zero corresponding 
to the nodes of the autocatalytic set.
A larger PFe indicates a faster exponential growth driven by a higher connectivity in the core of the ACS.
Notice that, more in general, a matrix can have more than one PFe (and corresponding PFE) if it has more than one  
ACS (two ACS are distinct if there is no path connecting the two cores).

In what follows, we show how this very simple model can both i) give us a novel understanding of the innovation system as
a process of cumulative causation, and ii) identify the core technologies in the evolving technological landscape.

\section{Autocatalytic structure of the technology network}\label{s:actnetworks}

\subsection{Mapping the network}

\begin{figure}
\centering
\includegraphics[height=.9\textheight]{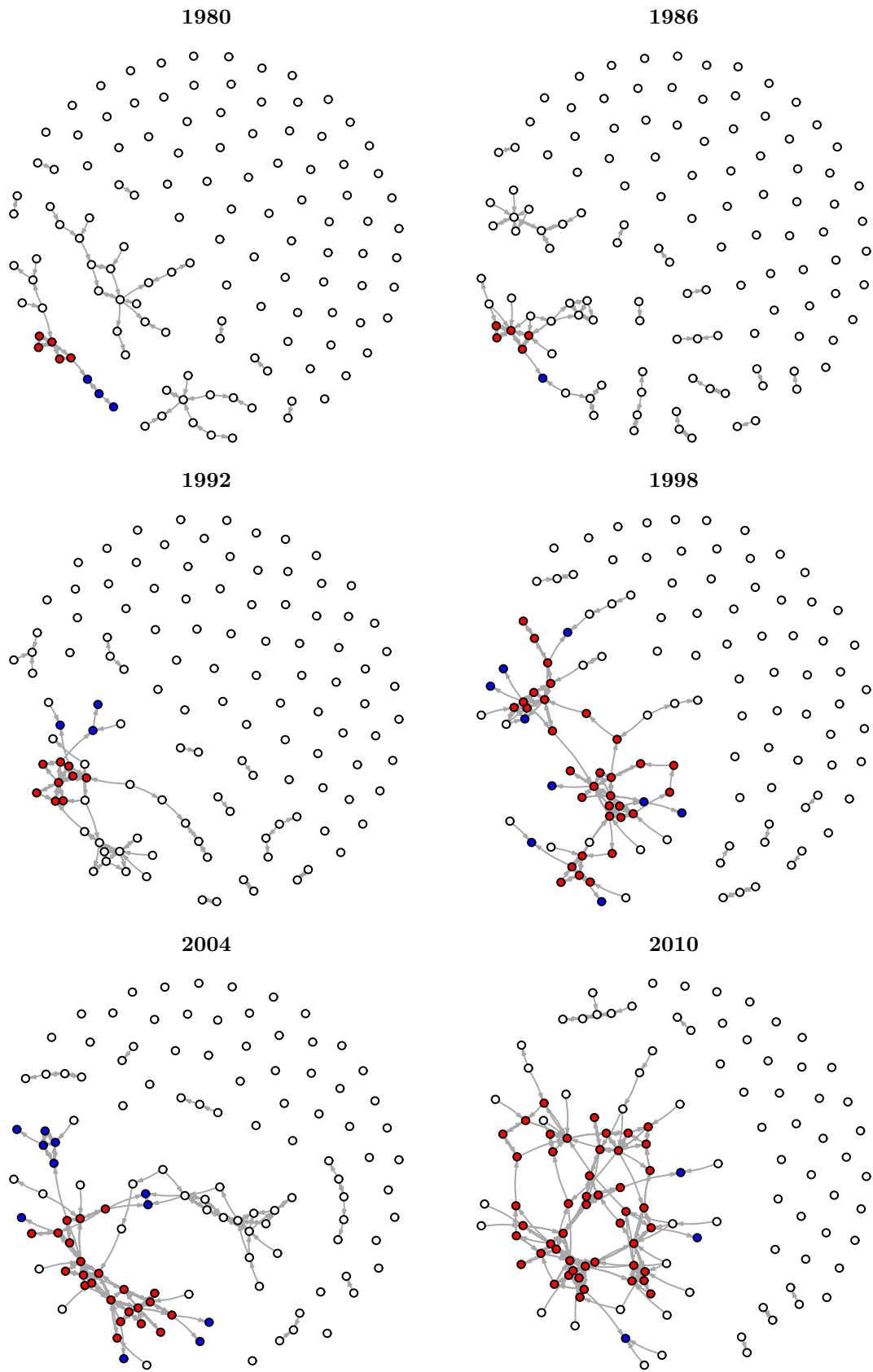}
\caption{The technology network in six different years.
The nodes are technology classes (IPC classification).
Links account for statistically significant relatedness.
Red nodes represent the \emph{core} of the Autocatalytic Set (ACS) of the network, blue nodes are the \emph{periphery} of the ACS, white nodes are outside the ACS.}\label{f:network}
\end{figure}

Figure \ref{f:network} reports a selection of technological networks in which the nodes are the IPC classes (the interested reader is referred to Section \ref{s:subclass} in the appendix for the analysis of the network of IPC subclasses). 
In these plots, nodes are linked whenever there is a statistically significant relation between them.
After generating the network of technology classes, we map the network searching for an autocatalytic structure as defined in Section \ref{s:acnetworks}.
A unique ACS is present in the network between 1980 and 2010, the size of which increases over time.
At the beginning of the sample, multiple clusters are present that do not form an autocatalytic set because no closed path between them is present.
By 1998 most clusters have been `captured' by the ACS, which in 2010 spans approximatively half of the technology network.
In particular, the core occupies the largest portion of the ACS and peripheral nodes are only a minority.

\begin{figure}[!ht]
\centering
\includegraphics[height=0.4\textheight]{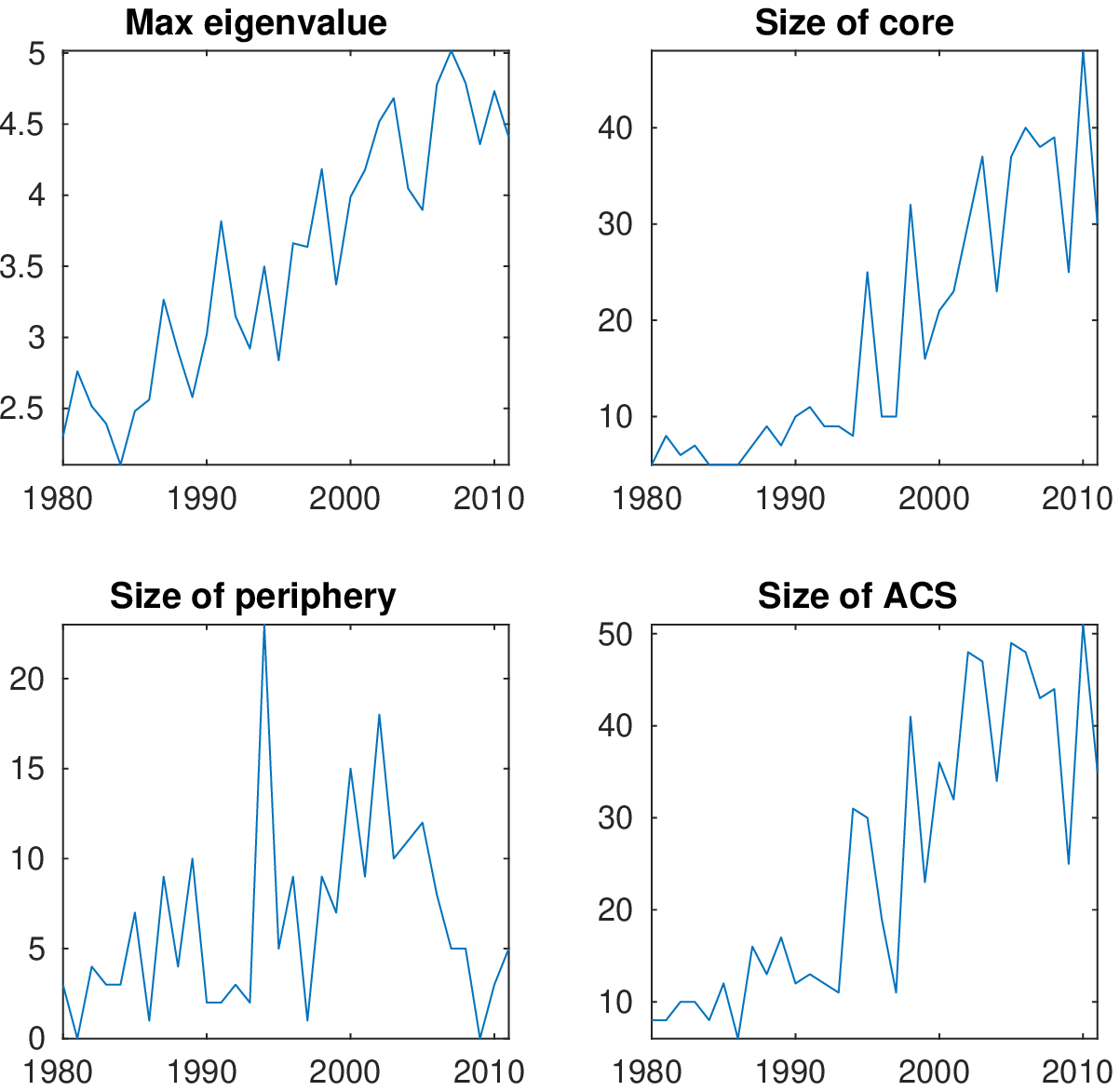}
\caption{Statistical indicators of the autocatalytic structure in the network of technology classes from 1980 until 2011.
\underline{Upper-left panel}: largest eigenvalue (Perron-Frobenius) of the network adjacency matrix.
\underline{Upper-right panel}: size of the core.
\underline{Lower-left panel}: size of the periphery.
\underline{Lower-right panel}: size of the ACS.}\label{f:acs}
\end{figure}

Figure \ref{f:acs} reports the time evolution of the relevant statistical indicators of the autocatalytic network configuration, namely the maximum eigenvalue (upper-left panel), the size (number of classes) of the core (upper-right panel), the size of the periphery of the ACS (lower-left panel) and the size of the entire ACS (lower-right panel). 
The plots clearly show that the autocatalytic character of the network becomes stronger in time and that its growth seems to occur in two phases: a first phase in which the periphery expands (bottom left panel) and a second one where most of technologies in the ACS transition to the core (see bottom left and bottom right panels).

\subsection{Fitness}\label{s:class_fitness}

Having observed the emergence of an autocatalytic structure in the technology network of classes, we want to understand how this structure affects the \emph{fitness}\footnote{
  Note that we employ fitness in its original biological interpretation, \emph{i.e.} the ability of a population to reproduce. 
  There is no relationship between fitness as we use it in this paper and the use of the term in the related literature of Economic Complexity \citep{tacchella2012new}.}
of the technological fields. 
In this article, we define fitness of a technological field the number of patent applications filed in a certain year featuring said field.
The idea is that a higher patenting rate is indicative of a higher technological productivity, and a proxy of the `innovativeness' of a technology class.
By comparing class fitness in different parts of the network -- namely the core, the ACS, and the rest of the nodes -- we test our hypothesis that technology classes inside the ACS display higher fitness than other classes.

\begin{figure}
  \centering
  \begin{subfigure}[t]{0.45\textwidth}
    \centering
    \includegraphics[height=0.18\textheight]{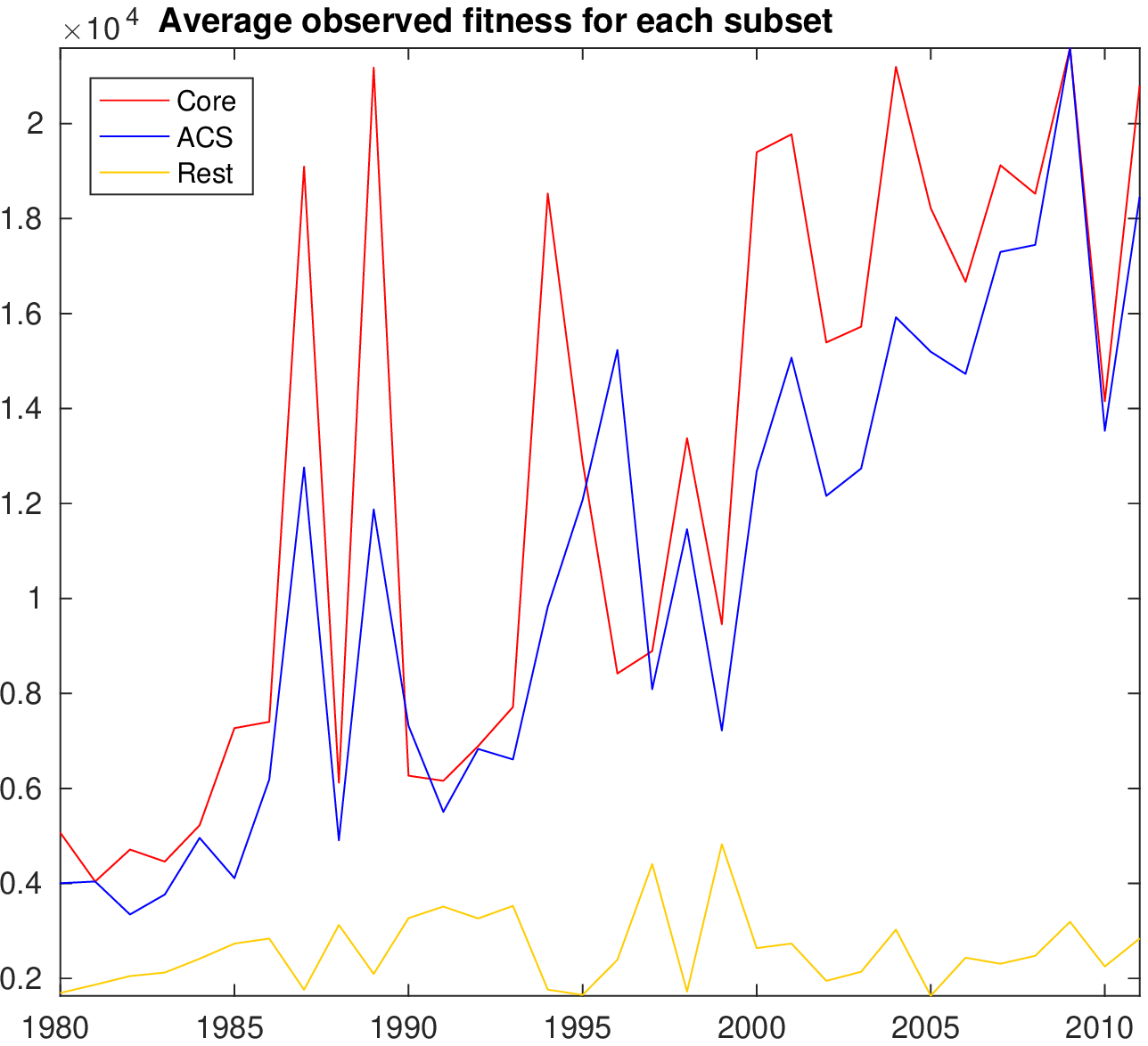}
    \caption{Average fitness.}\label{f:avgofit}
  \end{subfigure}
  \quad
  \begin{subfigure}[t]{0.45\textwidth}
    \centering
    \includegraphics[height=0.18\textheight]{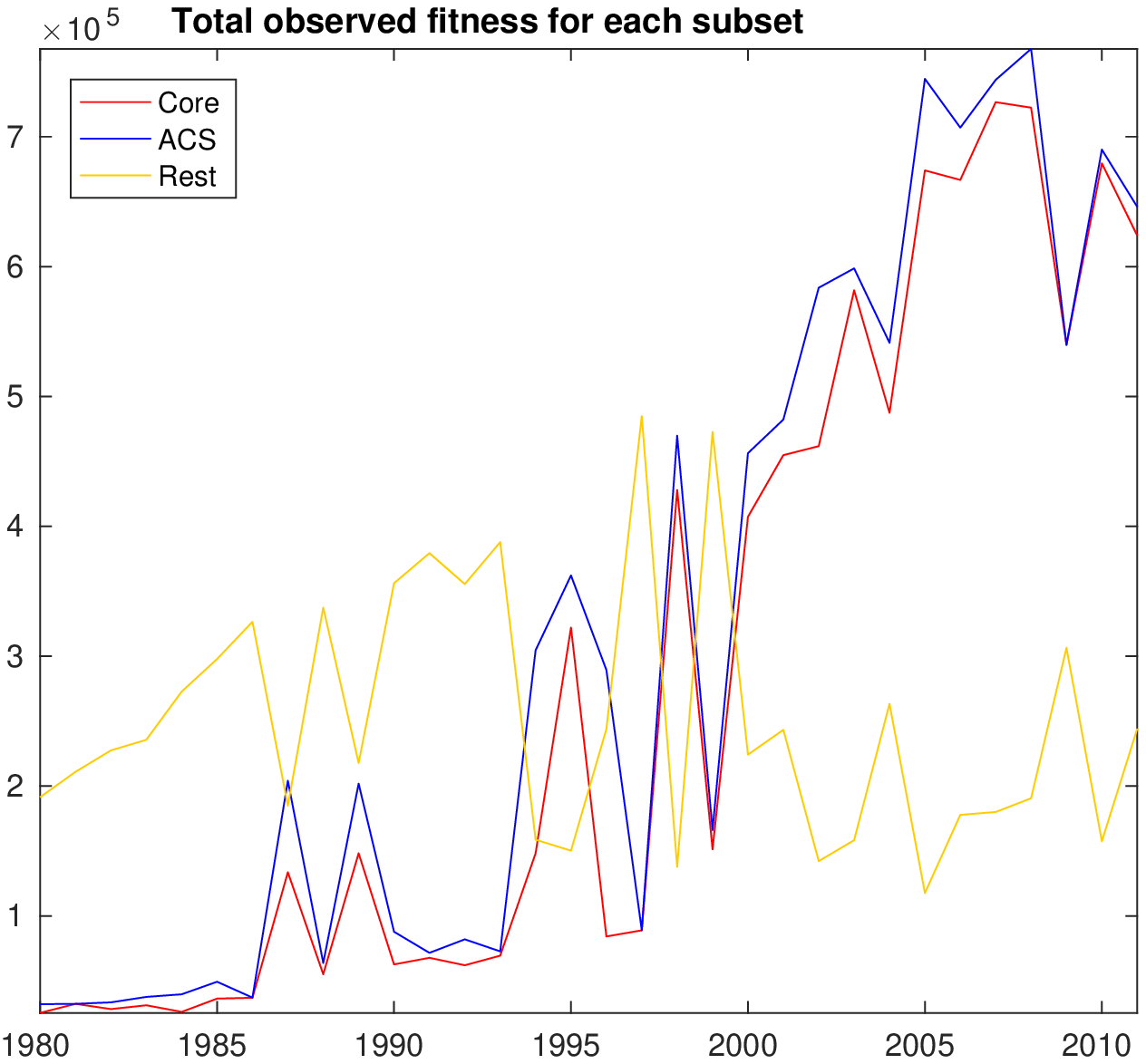}
    \caption{Total fitness.}\label{f:avgofit2}
  \end{subfigure}
  \quad
  \\
  \bigskip
  \begin{subfigure}[t]{0.45\textwidth}
    \centering
    \includegraphics[height=0.18\textheight]{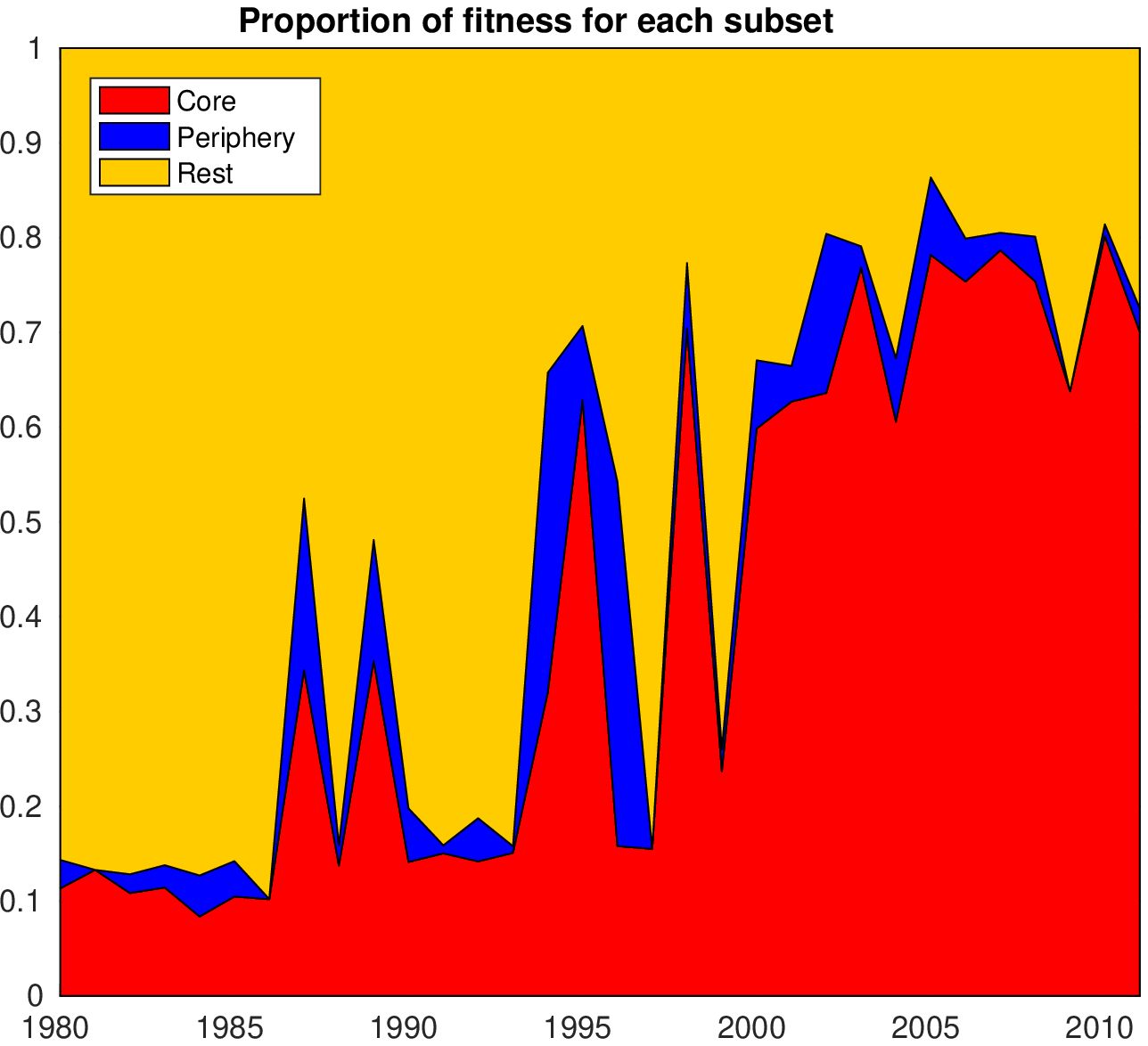}
    \caption{Share of total fitness.}\label{f:shares}
  \end{subfigure}
  \begin{subfigure}[t]{0.45\textwidth}
    \centering
    \includegraphics[height=0.18\textheight]{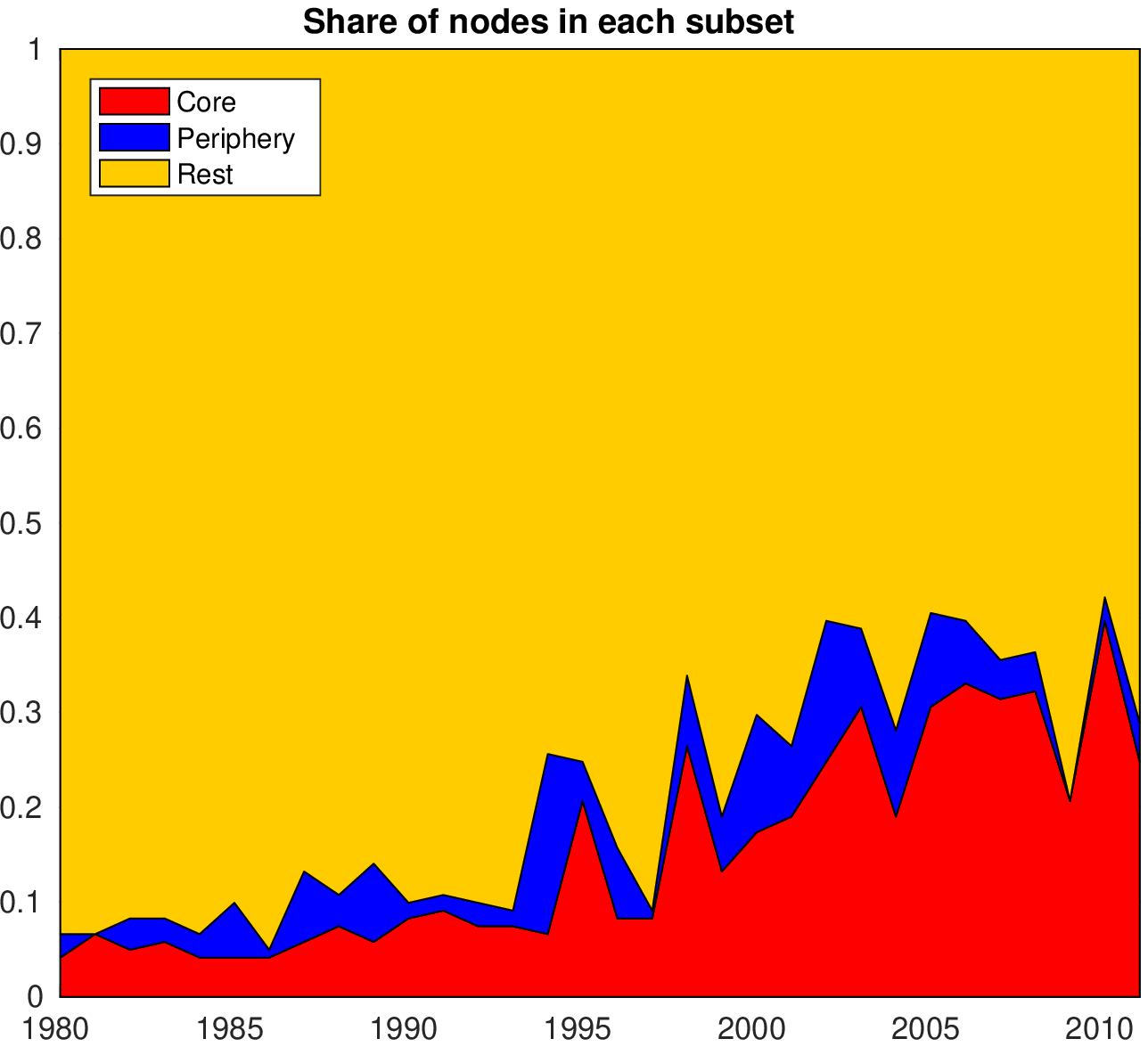}
    \caption{Share of nodes.}\label{f:shares_links}
  \end{subfigure}
  \caption{Total and average fitness, \emph{i.e.} total and average number of patents per IPC class in different portions of the technology network (ACS, core, and the rest of the network).}\label{f:avgofit_panels}
\end{figure}

Figure \ref{f:avgofit_panels} shows the time series of the fitness of the classes belonging to the core (red), to the whole ACS (blue), and to the rest of the technology network.
Figure \ref{f:avgofit} reports the average fitness of nodes in each of the above mentioned subsets.
It clearly emerges that the nodes in the ACS, and especially those in the core, show an increasing average fitness, while the fitness of the nodes in the rest of the network fluctuates around a constant value.
This value is twice smaller than the fitness of the nodes in the ACS at the beginning of the sample and is ten times smaller by 2011.
The average fitness in the ACS and in its core is largely of the same order of magnitude, although for nodes of the core it is almost always larger (it is lower only in 1990 and in 1996).
These figures lend support to the hypothesis that technology classes belonging to the ACS benefit from an autocatalytic advantage.
Figure \ref{f:avgofit2} proxies an absolute measure of fitness by reporting the time series of the total number of patents in the ACS, its core, and the rest of the network.
The plot shows that there were many more patents outside the ACS until the early 1990's, when only a few nodes were part of the ACS (Figure \ref{f:network}).
However, as soon as the size of the ACS started increasing substantially, this measure of absolute fitness quickly surpassed the total fitness of the rest of the network by a large measure.
Although less meaningful than the average fitness time series, this figure is evidence of the transition experienced by the network of technological fields, with the emergence of a large autocatalytic structure around the end of the century.
Our analysis suggests that technology classes belonging to the ACS are more innovative due to the positive effect of catalytic links on knowledge flows.
In particular, classes in the core of the ACS experience the positive feedback of self-reinforcing cyclical catalytic structures, which give rise to cumulative processes of innovation.
Figure \ref{f:shares} reinforces the above intuition by showing the evolution of the distribution of total fitness between the core of the ACS, the periphery, and the rest of the nodes.
It shows that in the early 1980's the autocatalytic structure was rather marginal in the network of technology classes, since it included only around 10\% of the total number of patents.
After 1985, an increasing trend started leading the ACS to gain weight in the network and become clearly predominant to the point that it concentrated almost 80\% of total fitness in the last decade of the sample in the face of a more contained increase in the share of nodes it includes, as shown in Figure \ref{f:shares_links}.
Considering that just above half of the classes in the network become part of the ACS in the same period, there is evidence of a strong correlation between the prominence of the ACS of the technology network and the intensity of patenting activity in the fields comprising it.

The above is a strong indication of the cumulative causation process behind innovation as identified by patenting activity.
If technology classes in the ACS grow while classes outside the ACS do not, then catalytic inter-linkages between classes are a relevant driver of innovation.
Taking patenting intensity as a meaningful proxy for innovation, the empirical evidence presented in Figure \ref{f:avgofit_panels} suggests that the substantial growth of a technology class is linked to its connection to other classes `feeding' into it with incoming links.
This is to say that a flow of knowledge stemming from the source of the directed link can provide new knowledge and thus the basis for a new patent in the target class.
The fact that stronger growth for a technology class comes from belonging to the core shows that innovation is fostered not only by technology spillovers from other classes, but especially from membership of a cycle of mutually reinforcing spillovers.

\subsection{Autocatalytic structure and database hierarchy}\label{s:autocal}

In this section, we turn to the hierarchical structure of the database to investigates its role in the autocatalytic structure of the technology network.
As mentioned in Section \ref{s:regsToFields}, PATSTAT adopts the International Patents Classification (IPC)\footnote{
Note in passing that the IPC is not the only existing classification of technological fields. However, the IPC is by far the most common and one of the few that allows to perform a broad international comparison.
}, 
which has a tree-like structure consisting of eight \emph{sections} at the root, which branch out into the progressively finer-grained \emph{classes}, \emph{subclasses}, and so on.
For the present analysis, we use classes and subclasses\footnote{The results for IPC subclasses are similar to those for classes; the interested reader is referred to  appendix \ref{s:subclass} for further details.} as nodes of the technology network, but we are also interested in understanding how the hierarchy induced by the IPC maps to the ACS.
To this end, we employ the IPC sections
\begin{itemize}
\setlength\itemsep{.05em}
\scriptsize
 \item HUMAN NECESSITIES (A)
 \item PERFORMING OPERATIONS; TRANSPORTING (B)
 \item CHEMISTRY; METALLURGY (C)
 \item TEXTILES; PAPER (D)
 \item FIXED CONSTRUCTIONS (E)
 \item MECHANICAL ENGINEERING; LIGHTING; HEATING; WEAPONS; BLASTING (F)
 \item PHYSICS (G)
 \item ELECTRICITY (H)
\end{itemize}
to cluster the nodes and see how sections map onto the structure of the technology network.
In particular, we ask if the ACS discriminates sections and if significant links cut across section borders.
These are not just questions of topological nature, because the answers shed light on whether recombinant innovation in general, and more 
specifically the cumulative causation process of autocatalytic structures take place mainly within classes or if, instead, it also involves broader connections between the coarse technological areas identified by IPC sections.

\begin{figure}
  \centering  
  \includegraphics[height=0.2\textheight]{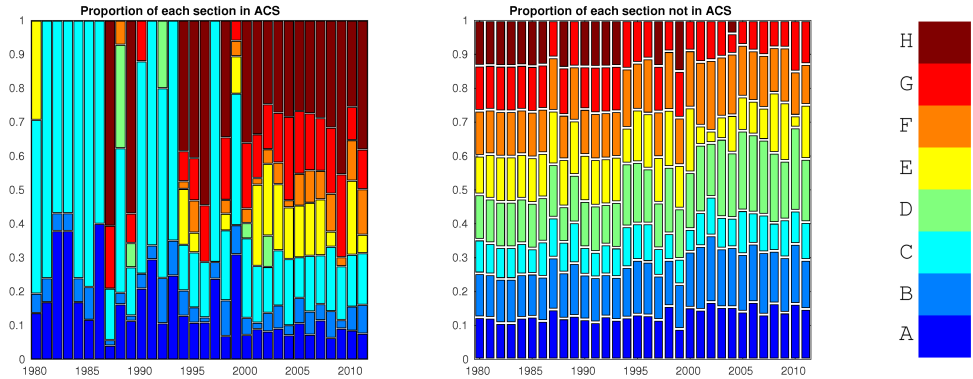}
\caption{Relative share of nodes in the ACS (left panel) and outside the ACS (right panel) for the eight technology sections of the database.}\label{f:propsec}
\end{figure}

Figure \ref{f:propsec} reports of the share of each IPC section within and outside the ACS thus showing how relevant each section is for the two subsets of nodes.
The main result, displayed in the left panel, is that the ACS is not at all static. 
At the beginning of the sample, the ACS consists almost entirely of the classes contained in section C (Chemistry, metallurgy), the only exception being one node of section A (Human necessity).
However, over time a more variegated picture unfolds and, by 2011, the ACS spans all sections but one -- D (Textiles; paper) -- which remains outside the ACS consistently and only makes a few sporadic appearances.
A further observation suggested by Figure \ref{f:propsec} concerns the different composition of the ACS and the rest of the network; while the latter presents a quite uniform distribution of sections along the whole period 1980-2011, the ACS is characterised by less uniformity and a richer dynamics.

Concerning the above non-uniformity, it is also possible to quantitatively assess the non-uniform distribution of sections in the growing autocatalytic structure of the technology network.
To this end, we can imagine having an urn filled with marbles of different colours, each corresponding to a distinct section, in which every marble represents an IPC class. 
There is a different number of marbles of each colour in the urn and we want to sample as many marbles as there are classes in the ACS. 
The null hypothesis is that the sampling is random, meaning that marbles are sampled blindly from the urn. 
The alternative hypothesis is that the sampling process is preferential and tends to privilege a specific subset of colours.
We test the null hypothesis of random sampling against the alternative hypothesis of biased sampling from Fisher's non-central hypergeometric distribution \citep{Fog08}
and find that in every year from 1980 until 2011, the statistic is above the critical value for a  a significance of 5\% (see appendix \ref{s:evo_and_signif} for further details).
This indicates a significant bias in favour of some sections in terms of occupancy of the ACS over the period considered and shows that, though 
the expansion of the ACS brings more sections in the ACS and its core, a non-uniform distribution remains.

\begin{figure}[!ht]
\centering
\includegraphics[width=0.75\linewidth]{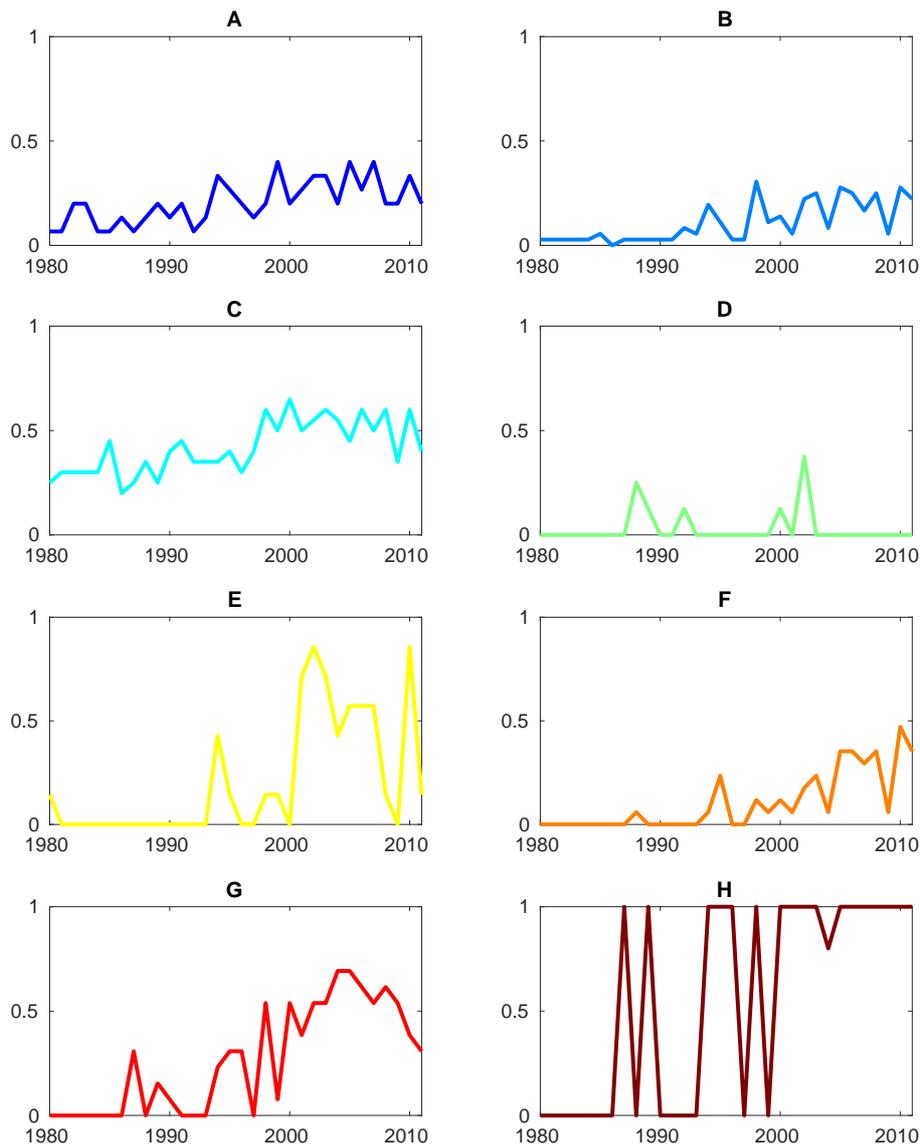}
\caption{Time series of the proportion of nodes in the ACS for each of the eight technology sections.
The plotted values are the ratio, for each section, of the number of classes belonging to the ACS to the total size of the section.}\label{f:propnodesep}
\end{figure}

The pattern observed in Figure \ref{f:propnodesep} is evident also if we look at the share of nodes of each section belonging to the ACS.
For instance, Figure \ref{f:propnodesep} shows that one section, H (Electricity), is entirely contained in the core of the ACS in the more recent years of the sample, while the situation of class E (Fixed construction) is less clear, also due to large fluctuations in the final decade.
Sections F and G seem to display a somewhat growing trend in the share of classes the y contribute to the ACS.
The remaining sections, have a relatively stable share of classes in the ACS.
The data relative to the last decade highlights the strong transversal character of the autocatalytic structure of the technology network.

\begin{figure}
  \centering
  \begin{subfigure}[t]{0.43\textwidth}
    \centering
    \includegraphics[height=0.2\textheight]{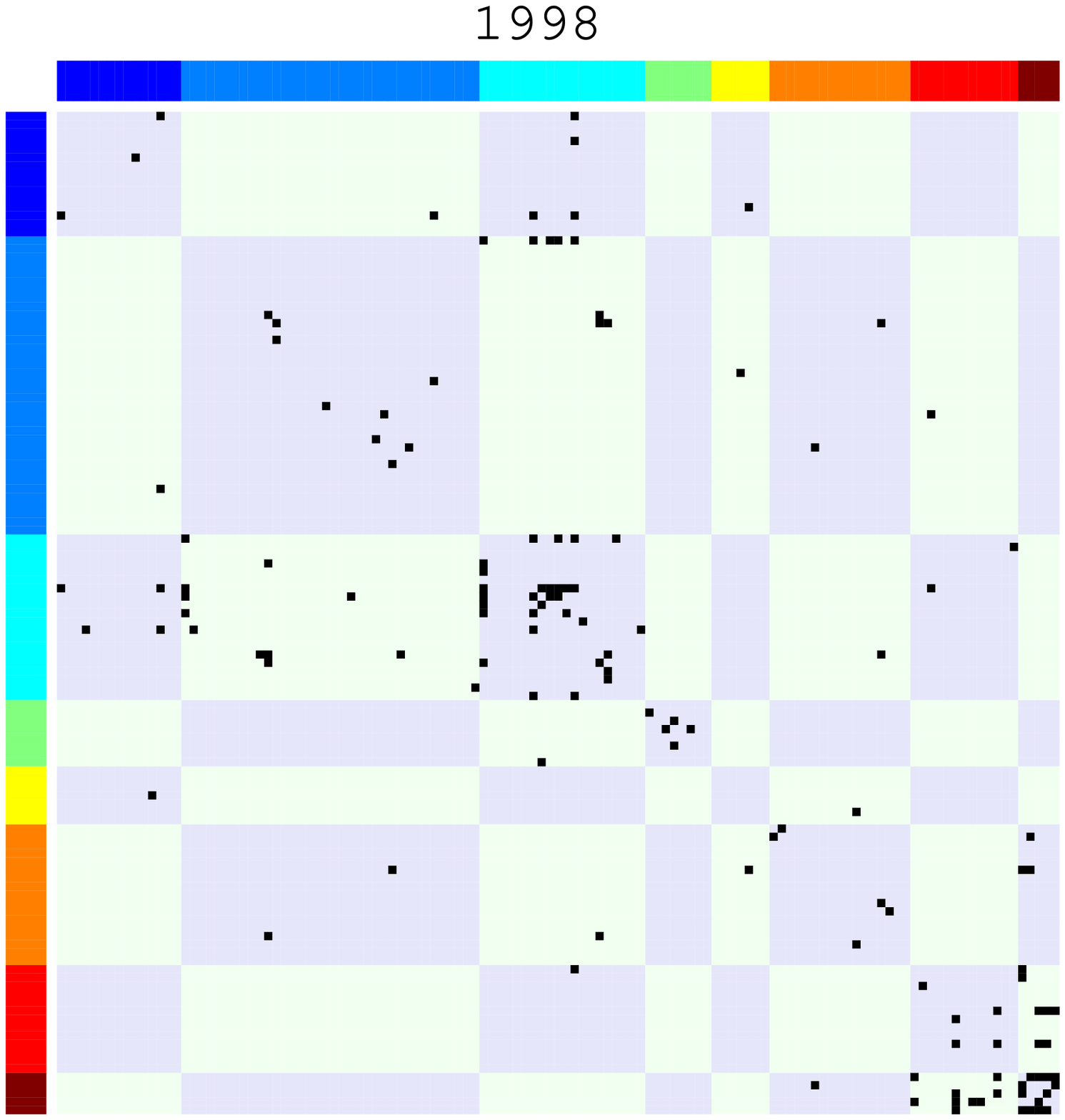}
    \caption{Technology network for 1998.}\label{f:heat_mat}
  \end{subfigure}
  \quad
  \begin{subfigure}[t]{0.42\textwidth}
    \centering
    \includegraphics[height=0.2\textheight]{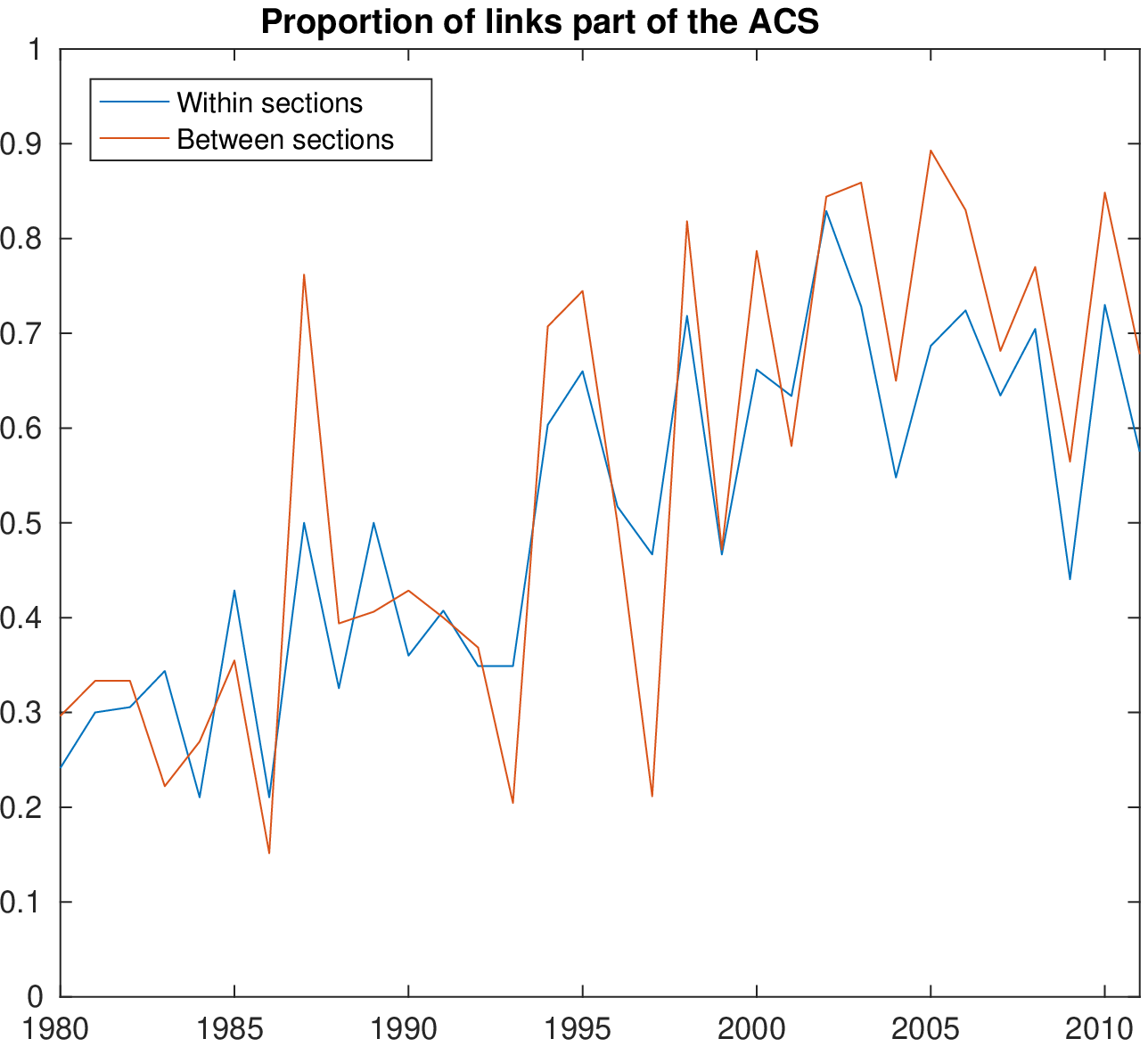}
    \caption{Within- and between-section links.}\label{f:heat_plot}
  \end{subfigure}
  \caption{In the left panel, the 121 IPC technology classes are ordered based on sections, which go from A (Human necessities -- dark blue -- in the top-left corner), to H (Electricity -- brown -- in the bottom right corner).
 Black elements represent statistically significant links between classes. The right panel plots the time series of within section and between section links over time.}\label{f:heat}
\end{figure}

Up to now we have seen that the expansion of the ACS has brought more sections in the ACS and its core, though a non-uniform distribution has remained, with some sections that are almost absent from the ACS throughout the whole period.
Moreover, there is a shift from A (Human necessities) and C (Chemistry and Metallurgy) towards H (Electricity) as the most important sections of the ACS, suggesting that there has been a transition from Chemistry and Metallurgy in the 1980's to Electricity in the $21^{st}$ century as driving forces of innovation.
The relevance of individual sections can be further clarified by looking at the adjacency matrix of the technology network displayed in Figure \ref{f:heat_mat} and the relative distribution of links within and between sections plotted in Figure \ref{f:heat_plot}.
Notice that the matrix of Figure \ref{f:heat_mat} is sorted according to a lexicographic ordering of the IPC class codes, so that nodes are grouped together based on the section they belong to.
The presence of a block-diagonal structure would indicate that most links occur between classes of the same section.
However, evidence of such a pattern appears to be mixed both in Figure \ref{f:heat_mat}, which depicts the adjacency matrix of the technology network for 1998 (the central year of the sample\footnote{
 The interested reader is referred to Section \ref{s:adj_robust} information for a depiction of the adjacency matrix in a broader selection of years.
 }) 
and Figure \ref{f:heat_plot}, which summarizes the evidence for the entire period.
Note that sections in the technology network are not all equal in terms of \emph{within} and \emph{between} connectivity.
For example, sections B and D seem to have mostly self-contained significant links, suggesting that the hierarchical structure of the IPC captures the extent of knowledge spillovers for technologies related to transporting, textiles, and paper.
On the other hand, sections C and A, and likewise sections G and H, appear to share many links that cut the section border.
This suggests that relevant spillovers can take place between `distant` (and versatile) technologies and that a relevant role is played by the subsets of the autocatalytic structures in affecting the distribution of links.
In other words, innovation, as measured by the production of patents, can potentially spawn as many connections between sections as it does within sections.
The autocatalytic structure thus has at least comparable importance to that of sections in defining the boundaries of the drivers of technological progress.

\section{Conclusions}\label{s:conclusion}

This study is a first step to uncover the cumulative causation processes driving technological change by detecting an autocatalytic structure of patent databases.
Our results can be summarized in three main points.
First, the technology landscape described by the network of patent technology codes is characterized by a clear autocatalytic structure that has grown over the years to encompass most technology classes. 
Second, the classes that are involved in the autocatalytic set perform better in terms of innovativeness as measured by the rate of growth in the number of patents containing them. 
Finally, the autocatalytic structure of the technology network is more evident and stronger than the hierarchical structure of the database, in that as many links connect classes from different sections as they do classes from the same section. 
This implies that recombinant innovation arising from interdisciplinary technological interactions is a stylized fact of technological change.

We believe that our approach based on detecting autocatalytic structures can be successfully extended in more fundamental ways.
We owe at least part of the inspiration to study autocatalysis in technological systems to the work of \cite{kauffman1986autocatalytic,Jain-Krishna:2001, Jain-Krishna:2002}, who proposed autocatalytic networks as a model of self-organization for interlinked biological species. 
However, a distinctive characteristic of those models is the interplay of a \emph{fast dynamics} driving species evolution and a \emph{slow dynamics} `reshuffling' the links of the network through species replacement or mutation. 
In this work, we instead assume that the equivalent of the fast dynamics acts on the population of technological codes and observe the changing network without providing a model for its evolution or its relation to the slow dynamics.
A reason for this is that, while it is reasonable to assume that population and network evolution take place at very different time scales in biological networks, this seems less plausible in the domain of innovation, where intuition suggests that success of individual technological fields and their mutual interactions might change at more similar speeds.

In principle technological codes are a powerful device to explore not only the different time scales of the dynamics shaping technological progress, but also the different scales of technological definition and categorization. 
For example, recent studies have observed that about 60\% of new patents use novel combinations of codes \citep{Youn-etal:2015} taken from the most recent version of the technological classification.
Moreover, an interesting property of codes is the fact that the classifications they are drawn from are not static, but rather change over time to keep up with the pace of technological change.
In fact, recombination of existing knowledge appears as a distinguishing feature of innovation, a stylized fact that can be directly observed through changes of the classification system \citep{gkotsis2016technological, lafond2017long}.
We envisage two main avenues for future research stemming from our study: first, an empirical analysis of autocatalytic sets at different scales of technological classification, to uncover possible fractal structures; second, a modelling framework that can reproduce and explain the statistical features of the empirical network of technology classes. 
From a methodological viewpoint, it would be interesting to explore the same questions using fundamentally different data about patents, such as citation networks \citep{hall2005market, erdi2013prediction} or co-occurrences of technology codes within patent documents or families.

This article is the initial phase of a broader research project aimed at empirically assessing and understanding cumulative causation in social systems \citep{Room-etal:2017}.
Within the realm of technological change, the results of this line of research bears potentially relevant implications for technological investment strategies at the corporate and institutional level as well as for innovation policy. 
More specifically, understanding the role of individual technological fields in the evolution of the wider technology system can be of the utmost importance for designing policies that address the different challenges of present times, from economic development and inequality to energy, security, and climate change.

% 
% \vfill
% \paragraph{Data Accessibility}Data, codes, and detailed explanation of their structure, have been submitted to dryad. Instructions for download will be available at \url{http://www.bath.ac.uk/imi/pdf/DCICSS-2017-3.pdf} in case of manuscript acceptance.
% \paragraph{Author Contributions}EP, PZ and GR designed the study; LN and EP prepared the data sample; EE analysed the data; LN, EP and PZ interpreted the results and wrote the manuscript. All authors gave final approval for publication.
% \paragraph{Competing Interests}We declare we have no competing interests.
% \paragraph{Funding}EP and LN acknowledge financial support from the Italian Project of National Interest CRISISLAB (MIUR). GR, PZ and EE acknowledge financial support from the Institute of Mathematical Innovation (IMI) of the University of Bath.
% \paragraph{Acknowledgements}We are very much indebted to colleagues in the DCICSS (Dynamic of Cumulative Innovation in Complex Social Systems) team for their contribution to the conceptual framework on which the research presented in this article resides. 
% We are also grateful to the anonymous referees for their comments, which have helped us improve the quality of our manuscript.
% Of course, all mistakes and flaws remain solely ours.
% Further information about the DCICSS project can be found at \\ \url{http://www.bath.ac.uk/imi/research/DCICSS-project.html}.
% 

\pagebreak
\bibliographystyle{vancouver}
\bibliography{bibliography_acn.bib}

\pagebreak
\appendix

\section{Results for IPC subclasses}\label{s:subclass}

\begin{figure}[!ht]
\centering
\includegraphics[height=0.35\textheight]{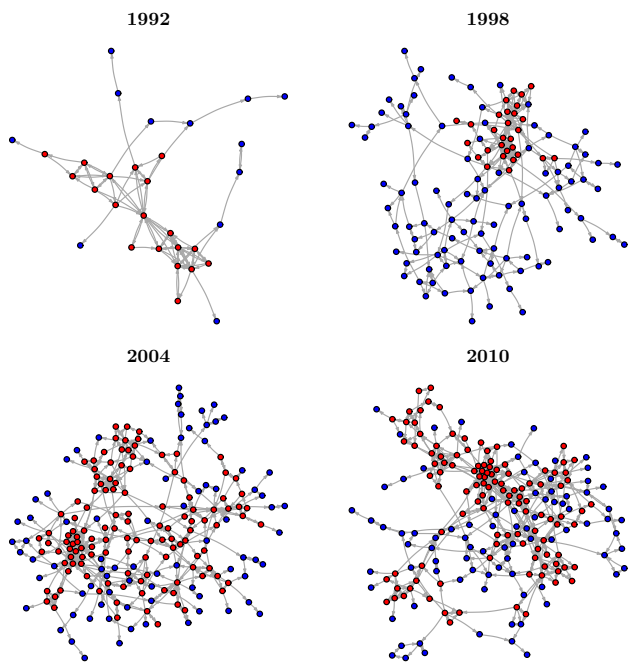}
\caption{Network of technology subclasses: snapshots of the ACS.}\label{f:zoomed_acs}
\end{figure}

\begin{figure}
  \centering
  \begin{subfigure}[t]{0.4\textwidth}
    \centering
    \includegraphics[height=0.18\textheight]{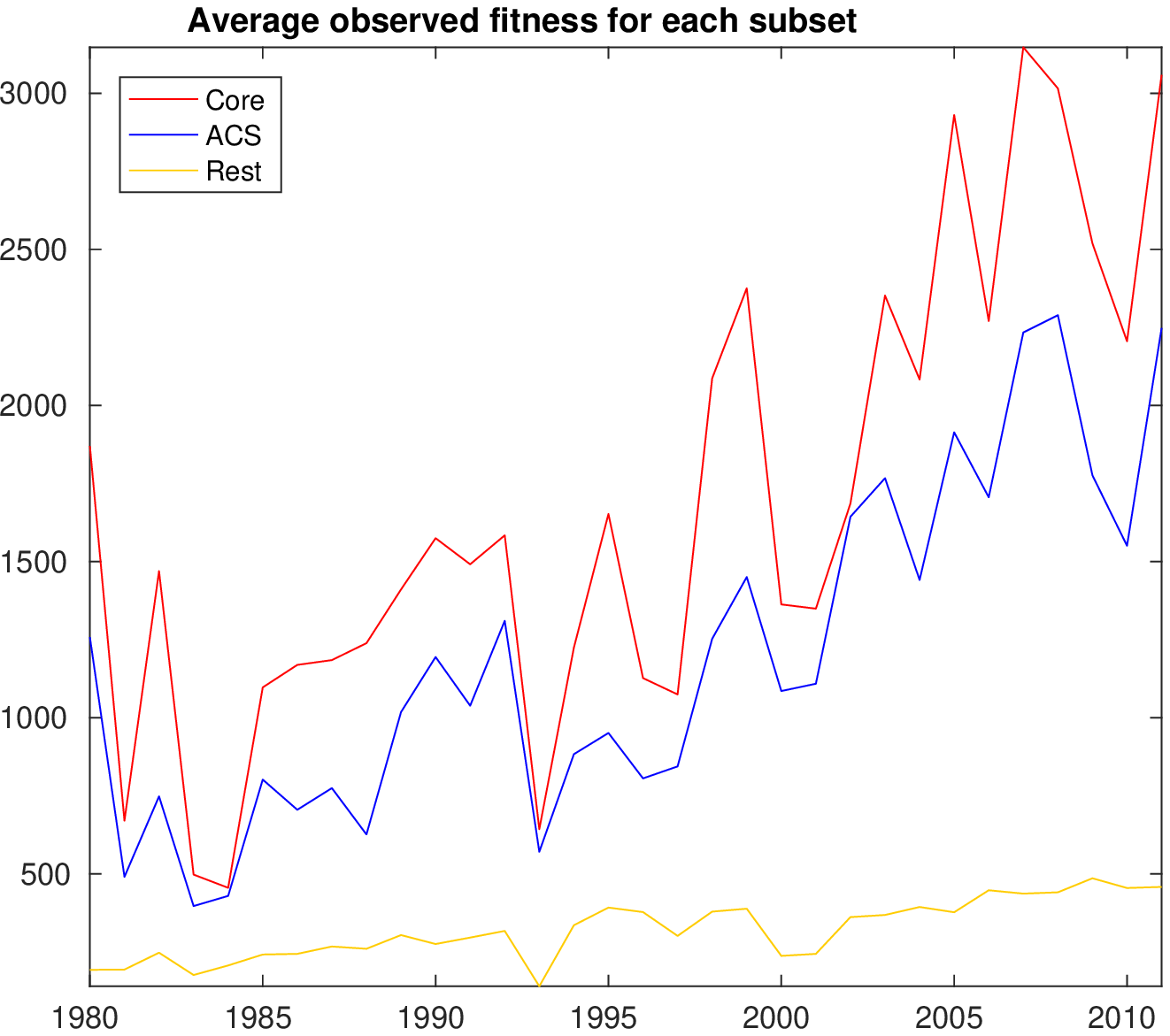}
    \caption{Average fitness.}\label{f:zoomed_avgofit}
  \end{subfigure}
  \quad
  \begin{subfigure}[t]{0.4\textwidth}
    \centering
    \includegraphics[height=0.18\textheight]{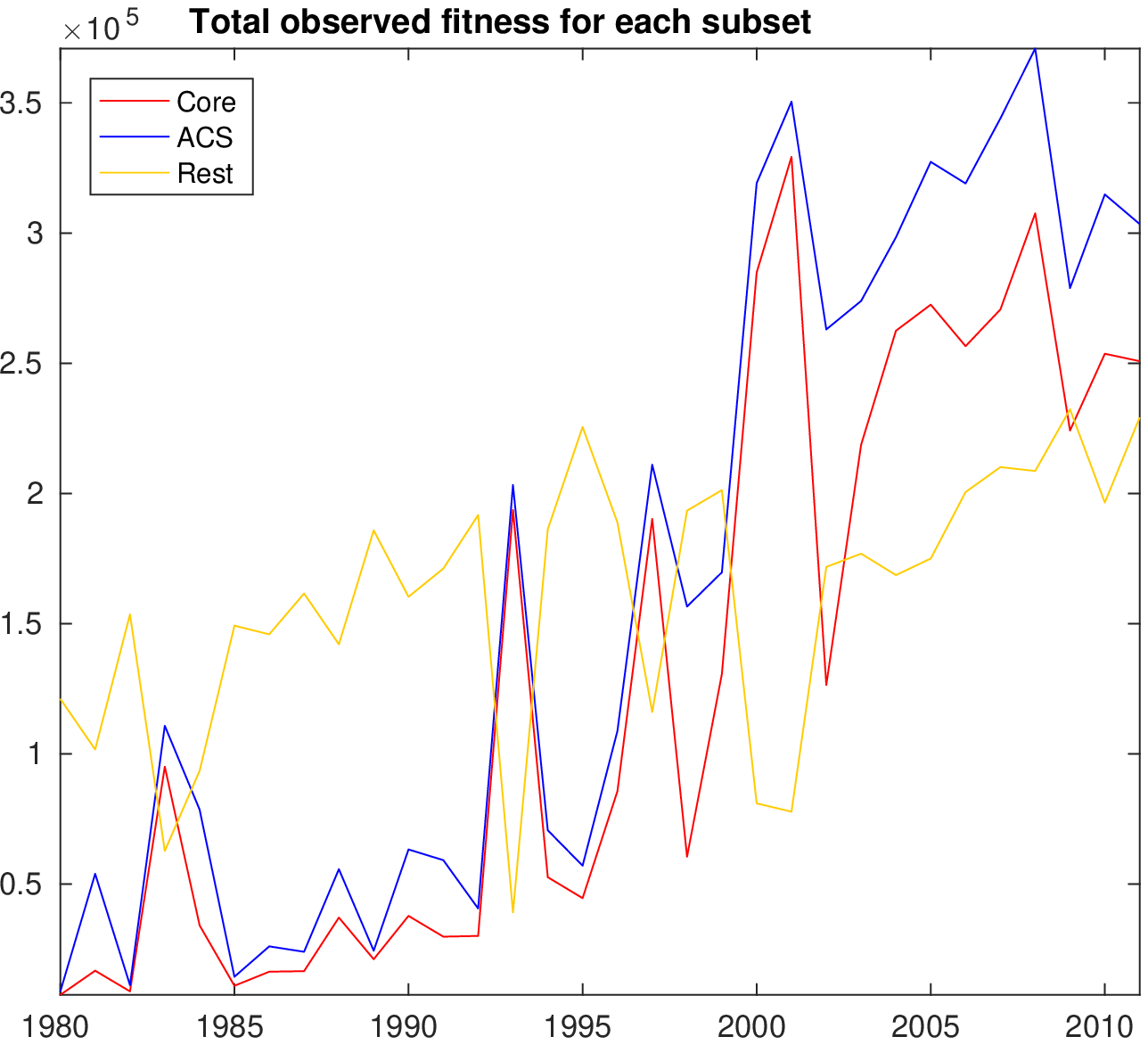}
    \caption{Total fitness.}\label{f:zoomed_avgofit2}
  \end{subfigure}
  \\
  \bigskip
  \begin{subfigure}[t]{0.4\textwidth}
    \centering
    \includegraphics[height=0.18\textheight]{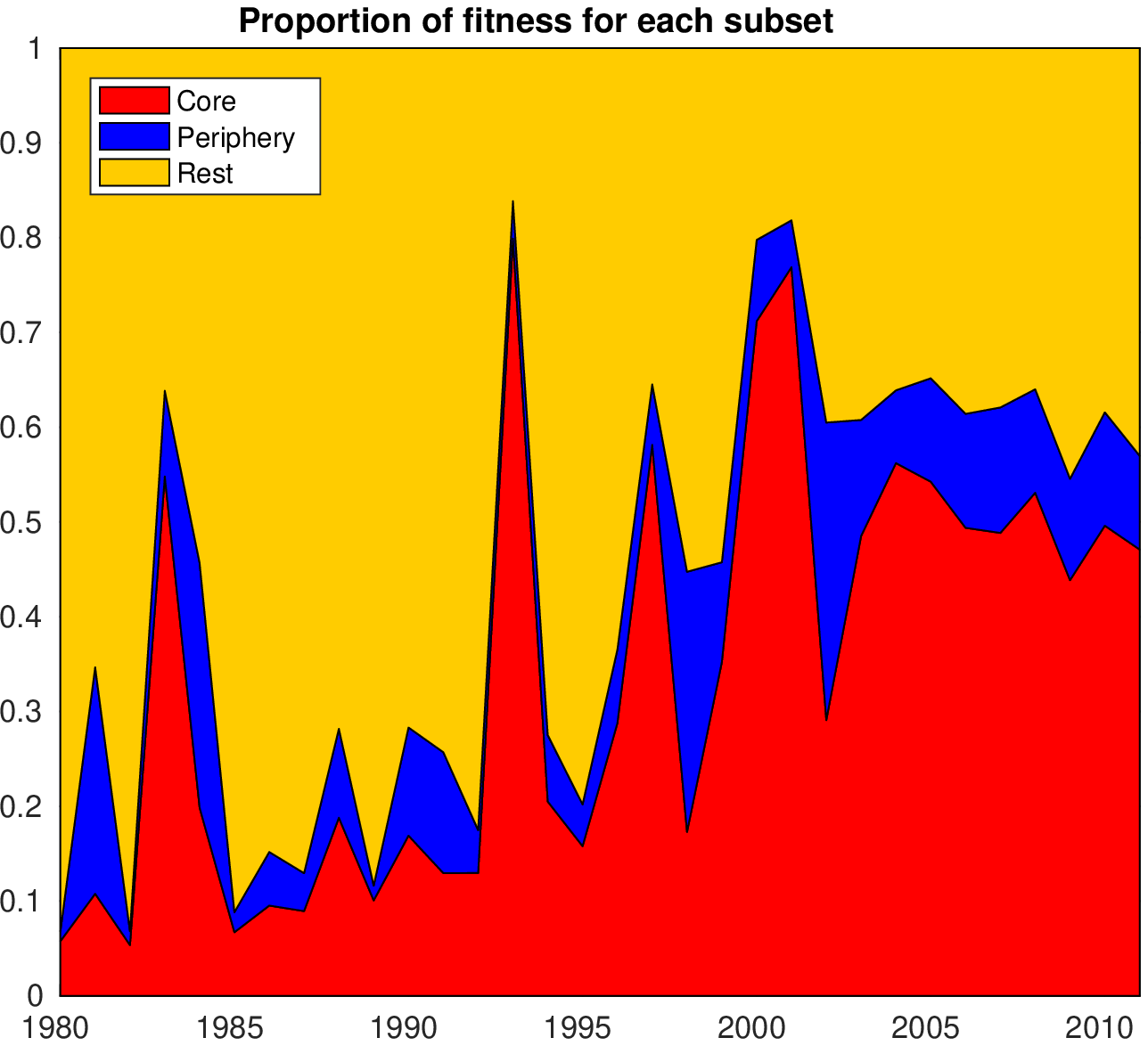}
    \caption{Share of total fitness.}\label{f:zoomed_shares}
  \end{subfigure}
  \begin{subfigure}[t]{0.4\textwidth}
    \centering
    \includegraphics[height=0.18\textheight]{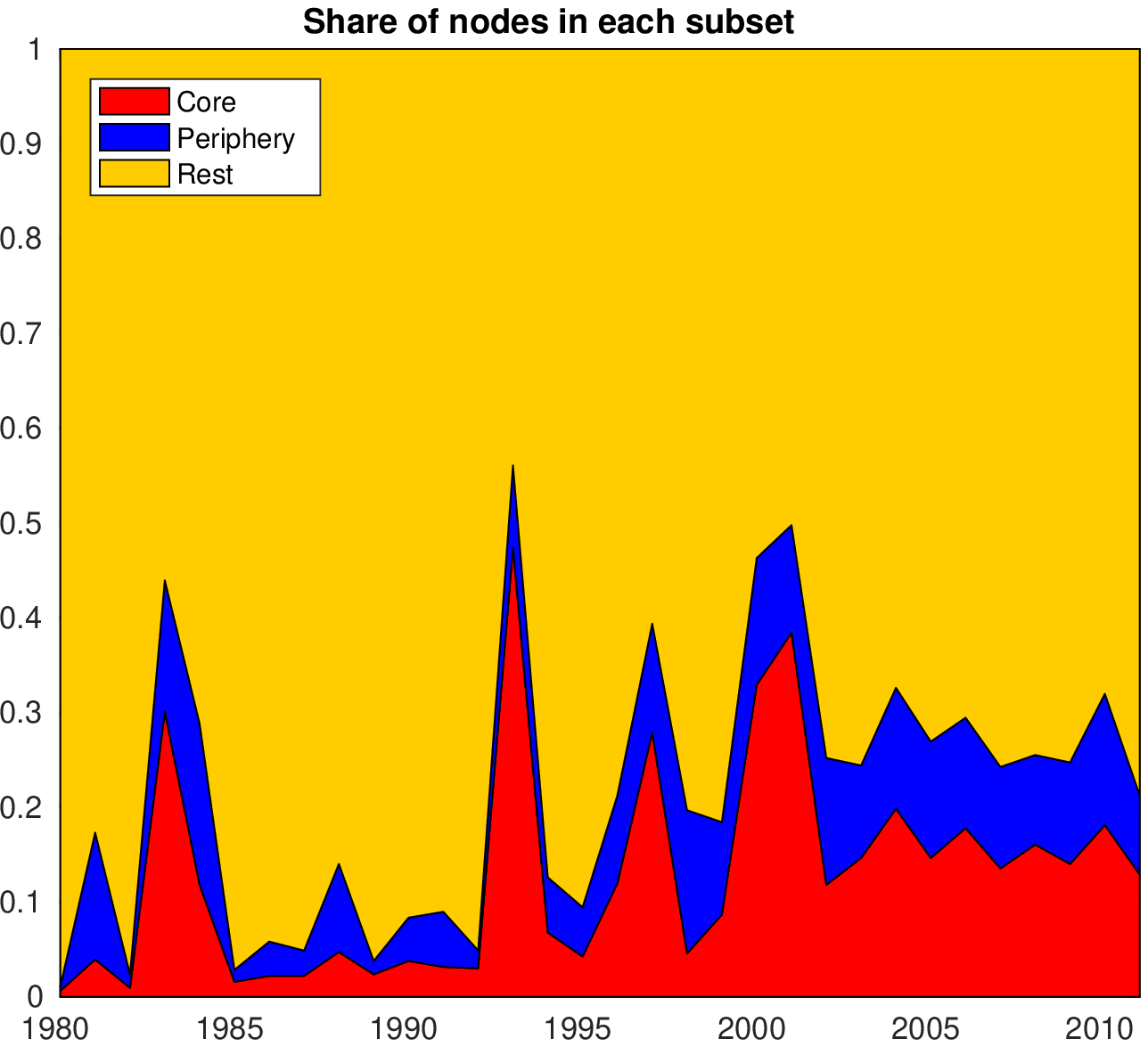}
    \caption{Share of nodes.}\label{f:zoomed_sharesLinks}
  \end{subfigure}
  \caption{Total and average fitness, \emph{i.e.} total and average number of patents per subclass in different portions of the technology network (ACS, core, and rest).}\label{f:zoomed_avgofit_panels}
\end{figure}

In this section, we examine the network of technologies defined by IPC subclasses.
Figure \ref{f:zoomed_acs} displays a selection of snapshots of the ACS at the subclass level, while Figures \ref{f:zoomed_avgofit} and \ref{f:zoomed_avgofit2} respectively show the time series of the average fitness and the total fitness of subclasses in the ACS, its core and the rest of the network.
As with IPC classes (see Figure \ref{f:avgofit_panels}), we find a clear positive trend of fitness for the ACS and its core over time.
In particular, the average fitness of subclasses in these two sets grows from about 1000 patents per subclass in the ACS to about 2000 patents per subclass, and the average fitness for the core increases even more, to about 3000 patents per subclass (Figure \ref{f:zoomed_avgofit}).
At the same time, the average fitness in the rest of the network only barely increases reaching around 500 patents per subclass.
Also the figures for the total fitness are indicative (Figure \ref{f:zoomed_avgofit2}) and show a lower level of fitness for ACS and core compared to the rest of the network in the first two decades of the investigated time period.
The rest of the network is later surpassed by the fast growing ACS around the turn of the century.
If we discount for the fact that subclasses are still less numerous in the ACS (and we normalise by the total number of subclasses in a set) we obtain the picture presented by the average fitness in Figure \ref{f:zoomed_avgofit}.

Figure \ref{f:zoomed_shares} shows the relative share of the three subsets of the subclasses network in terms of absolute fitness (number of patents).
This figure presents a similar trend to the fitness in the network of IPC classes (Figure \ref{f:shares}): the ACS increases in relative terms, while the rest of the network shrinks.
This time, the periphery of the ACS maintains a sizeable share (about 10\%).
In the last year (2011), the aggregate share of the ACS is about 60\% for subclasses, compared to about 80\% in the network of classes.
However, considering that for subclasses the aggregate number of nodes in the various Autocatalytic sets is roughly one third of the population, the autocatalytic structure emerges as a dominant pattern also in the network of subclasses.

The evidence of the analysis performed confirms our hypothesis at the level of subclasses, that autocatalytic structures are fostering innovation as measured by the number of filed patents.
Our hypothesis is thus verified also at the level of IPC subclasses as it is at the level of classes: like for technology classes, also the finer structure defined by subclasses clearly shows that technology fields benefit from the inclusion in the ACS.
This implies that the self-reinforcing process of cumulative innovation giving rise to the same quicker pace of technological progress that we observed in Section \ref{s:class_fitness} is detected also in a more disaggregated representation technology network.

\section{Variety in the ACS: evolution and significance}\label{s:evo_and_signif}

\begin{figure}[!ht]
\centering
\includegraphics[height=0.2\textheight]{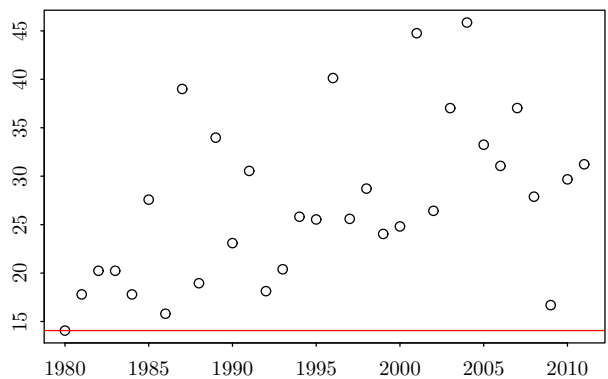}
\caption{Variety measure of sections with respect to autocatalytic structures.
Values are the log-likelihood ratio test statistic for each year contrasting the null hypothesis that occurrence in the ACS of each section happens with unbiased sampling without replacement for a fixed ACS size against the alternative hypothesis of biased sampling.
The horizontal red line corresponds to a 5\% significance level.}\label{f:variety}
\end{figure}

In this section, we investigate the composition of the ACS by running a statistical test of the uniformity of the contribution of IPC sections to the set.
The null hypothesis is that the occurrence of each section within the ACS is compatible with an unbiased sampling without replacement according to the generalised hypergeometric distribution. The alternative hypothesis is sampling from Fisher's non-central hypergeometric distribution \citep{Fog08}. 
The log-likelihood ratio statistic for each year is calculated and plotted in Figure \ref{f:variety}, with a lower value indicating evidence in favour of the null hypothesis. 
This statistic can be used as a measure of variety for the population of the ACS.
The asymptotic distribution of the log-likelihood ratio statistic is the chi-square distribution with 7 degrees of freedom. 
Its critical value corresponding to a 5\% significance level is 14.07. 
As shown in Figure \ref{f:variety}, in every year from 1980 until 2011 the statistics is above the threshold value, indicating a significant bias in terms of occupancy of the ACS by sections over the period considered, with an increasing trend.

\section{Links within and across IPC sections}\label{s:adj_robust}

\begin{figure}[!ht]
\centering
 \includegraphics[height=0.3\textheight]{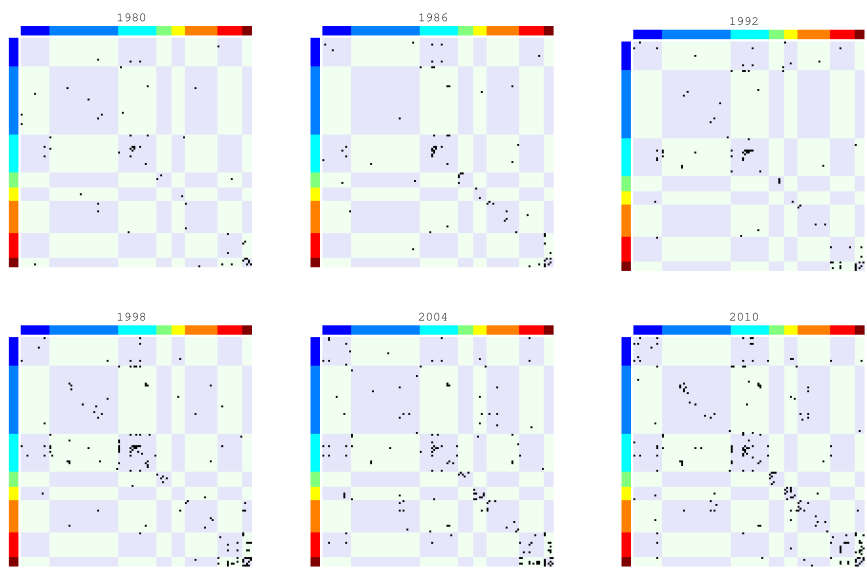}
\caption{Adjacency matrix of the network of technology classes for a selection of uniformly sampled years.
The 121 classes are ordered on the basis of their sections, which go from A (Human necessities -- dark blue -- in the top-left corner of each matrix), up to H (Electricity -- brown -- in the bottom right corner).
Black dots indicate that a link is present between two classes.}\label{f:heat_si}
\end{figure}

This section focuses on whether links in the ACS more frequently connect classes belonging to the same IPC section or to different sections.
Figure \ref{f:heat_si} shows a selection of snapshots of the technology network adjacency matrix sampling the analysed time period at regular intervals from 1980 to 2010. 
Comparing Figure \ref{f:heat_si} with Figure \ref{f:heat_mat} of Section \ref{s:autocal}, it clearly emerges that the structure of the matrix is robust over time and that the point made in the main text concerning the structure of links in the technology network is generally valid.
In particular, it can be seen that over time the number of links in the network increases but a significant proportion of them lies both within and across sections.
This indicates that important knowledge flows connect classes of different sections, and that recombinant innovation is a stylised feature of technological change.
Put differently, sections do not affect the distribution of links between classes alone, since between-section co-occurrences also appear to play a relevant role in some sections.

\end{document}